\documentclass[twocolumn,english,prb,aps,floats,showpacs]{revtex4}
\usepackage[T1]{fontenc}
\usepackage[latin1]{inputenc}
\usepackage{amsmath}
\usepackage{graphicx}
\usepackage{amssymb}

\makeatletter

\providecommand{\tabularnewline}{\\}

\usepackage{graphicx}
\usepackage{psfig}

\usepackage{babel}
\makeatother
\begin{document}

\title{Theory of nonlinear optical properties of phenyl-substituted polyacetylenes }

\author{Alok Shukla}

\affiliation{Physics Department, Indian Institute of Technology, Powai, Mumbai
400076 INDIA}

\begin{abstract}
In this paper we present a theoretical study of the third-order nonlinear
optical properties of poly(diphenyl)polyacetylene (PDPA) pertaining
to the third-harmonic-generation (THG) process. We study the aforesaid
process in PDPA's using both the independent electron H\"uckel model,
as well as correlated-electron Pariser-Parr-Pople (P-P-P) model. The
P-P-P model based calculations were performed using various configuration
interaction (CI) methods such as the the multi-reference-singles-doubles
CI (MRSDCI), and the quadruples-CI (QCI) methods, and the both longitudinal
and the transverse components of third-order susceptibilities were
computed. The H\"uckel model calculations were performed on oligo-PDPA's
containing up to fifty repeat units, while correlated calculations
were performed for oligomers containing up to ten unit cells. At all
levels of theory, the material exhibits highly anisotropic nonlinear
optical response, in keeping with its structural anisotropy. We argue
that the aforesaid anisotropy can be divided over two natural energy
scales: (a) the low-energy response is predominantly longitudinal
and is qualitatively similar to that of polyenes, while (b) the high-energy
response is mainly transverse, and is qualitatively similar to that
of \emph{trans}-stilbene.
\end{abstract}

\pacs{78.30.Jw,78.20.Bh,42.65-k}

\maketitle

\section{Introduction}

\label{intro}

Conjugated polymers are believed to be one of the most important materials
for the future nonlinear opto-electronic devices\cite{prasad}. The
nonlinear optical response of these materials originates mainly from
the $\pi$ electrons which (a) are delocalized along the backbone
of the polymer (b) while being completely localized as far as the
motion in a direction transverse to the backbone is concerned. It
is this one-electron picture of localized-delocalized $\pi$ electrons
which provides an intuitive understanding of the large response of
these materials to external fields. However, the role of electron-correlation
effects on the linear and nonlinear nonlinear optical response of
conjugated polymers has attracted considerable amount of theoretical
attention in recent years, and a number of conjugated polymers such
as \emph{trans}-polyacetylene, poly-(para)phenylene (PPP), poly-(para)phenylenevinylene
(PPV) etc. have been studied using the Coulomb correlated models in
recent works.~\cite{dixit,sumit,shakin,tretiak,yaron,shuai,aparna,lavren}
From several of these studies the picture that emerges is that the
nonlinear optical properties of various conjugated polymers for the
most part are determined by a small number of excited states. For
conjugated polymers whose structures are invariant under the inversion
operation (centro-symmetric) these states are: (a) $1B_{u}$, the
lowest-energy one-photon state, (b) $mA_{g}$, a two-photon state
higher in energy than $1B_{u}$ and with a strong dipole coupling
to it, and (c) $nB_{u}$, a one-photon state higher in energy than
$mA_{g}$ (but still within the conduction-band threshold), and with
strong dipole coupling to it.~\cite{dixit,sumit,yaron,aparna,lavren}
The question is whether this so-called {}``essential-states picture''
is peculiar to a few conjugated polymers, or is it universally shared
by all of them?

Recently, optical properties of a new class of conjugated polymers
called phenyl-disubstituted polyacetylenes (PDPA's)---which are obtained
by substituting the side H atoms of \emph{trans}-polyacetylene by
phenyl derivatives---have been the subject of several experimental
studies.\cite{tada1,tada2,liess,fujii1,gontia,sun,hidayat,fujii2}
The interesting aspect of the linear optical properties of these materials
is that they exhibit strong photoluminescence (PL), despite their
structural similarities to \emph{trans}-polyacetylene, which is well-known
to be nonphotoluminescent.\cite{gontia} In a series of theoretical
studies, we explained the strong PL exhibited by PDPA's in terms of
reduced electron-correlation effects caused by the delocalization
of exciton in the transverse direction because of the presence of
phenyl rings in that direction.\cite{shukla1,shukla2,shukla3} We
also showed that delocalization of exciton also leads to reduction
in the optical gaps of these materials as compared to \emph{trans}-polyacetylene.\cite{shukla1,shukla2,shukla3}
Moreover, we predicted that the delocalization of excitons will also
leave its signatures in form of a significant presence of transverse
polarization in the photons emitted during the PL in PDPA's, a prediction,
which since then, has been verified in oriented thin-film based experiments.\cite{fujii2}

In the present work, we have undertaken a systematic theoretical study
of nonlinear optical properties of PDPA's corresponding to the third-harmonic
generation (THG) process. There are several motivations behind the
present study, the first of which is studying those excited states
of PDPA's which will not be visible in the linear optics. In the THG
spectrum of centrosymmetric materials such as PDPA's, $A_{g}$-type
states appear as two-photon resonances, while the $B_{u}$-type states
appear as three-photon resonances, thus making the simultaneous exploration
of both the symmetry manifolds possible. The next motivation is to
explore whether one can understand the nonlinear-optical properties
of complex polymers such as PDPA's, in simple terms as enunciated
in the essential-states mechanism mentioned above. In particular,
for \emph{trans-}polyacetylene, THG spectrum is predicted to have
two prominent peaks corresponding to the essential states $mA_{g}$
and $nB_{u}$ mentioned above, in addition to the $1B_{u}$ peak.\cite{dixit,sumit}
Therefore, questions arise: (a) How does the nonlinear optical response
of PDPA's compare to that of \emph{trans}-polyacetylene? (b) What
is the nature of the essential states contributing to the nonlinear
optical properties of PDPA's? (c) Will transverse polarization also
play a role in the nonlinear optical properties as it did in case
of their linear optical properties? The last, but not the least of
our motivations behind the present study is to investigate whether
one can use conventional computational methods such as configuration-interaction
(CI) approach also to study large unit cell materials such as PDPA's
(fourteen electrons per cell). This is important in light of the fact
that with so many novel and increasingly complex materials being discovered
by ``nano-technologists'' on a routine basis, there will be novel
challenges for anyone interested in obtaining their theoretical understanding.
Therefore, it is important to apply time-tested theoretical approaches
such as the configurations-interaction (CI) method to study these
materials to explore whether they can withstand the challenges posed
by systems of increasing complexity. We address all these issues by
performing calculations on oligo-PDPA's of the third-order nonlinear
optical susceptibilities corresponding to the THG process using both
the independent-particle H\"uckel model, as well as Coulomb-correlated
Pariser-Parr-Pople (P-P-P) model. The P-P-P model-based calculations
were performed using the CI methodology, and approaches such as multi-reference-singles-doubles
CI (MRSDCI), and quadruples-CI (QCI) were used to compute the THG
spectra. For the H\"uckel model calculations, we considered oligomers
of PDPA's containing up to fifty unit cells, while for P-P-P model
calculations up to ten unit cell oligomers were considered. We computed
susceptibility component where all photons involved were polarized
in the chain direction (longitudinal component), as well as the component
where the photons were polarized perpendicular to the chain (transverse
component). We see that the THG spectrum is clearly divided over two
distinct energy scales: (a) the low-energy response is primarily concentrated
in the longitudinal component and is qualitatively similar to that
of \emph{trans}-polyacetylene, while (b) the high-energy response
is concentrated mainly in the transverse component and is similar
to that of \emph{trans}-stilbene. Upon examining the orbitals involved
in these responses we conclude that it is the low-energy chain-like
molecular orbitals (MOs) close to the Fermi level which are active
in the longitudinal component, while the transitions among the chain-like
and high-energy phenyl-based orbitals give rise to the transverse
component of the response. 

The remainder of this paper is organized as follows. In section \ref{method}
we briefly describe the theoretical methodology used to perform the
calculations in the present work. Next in section \ref{results} we
present and discuss the calculated nonlinear optical susceptibilities
of oligo-PDPA's. Finally, in section \ref{conclusion} we summarize
our conclusions.

\section{Methodology}

\label{method}

\begin{figure}
\begin{center}\includegraphics[%
  width=4cm]{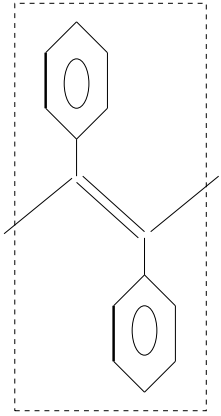}\end{center}

\caption{The unit cell of PDPA. The phenyl rings are rotated with respect
to the $y$-axis, which is transverse to the axis of the polyene backbone
($x$-axis)}

\label{fig-pdpa}
\end{figure}

The unit cell of PDPA oligomers considered in this work is presented
in Fig. \ref{fig-pdpa}. Ground state geometry of PDPA's, to the best
of our knowledge, is still unknown. However, from a chemical point
of view, it is intuitively clear that the steric hindrance would cause
a rotation of the side phenyl rings so that they would no longer be
coplanar with the polyene backbone of the polymer. The extent of this
rotation is also unknown, however, it is clear that the angle of rotation
has to be less than 90 degrees because that would effectively make
the corresponding hopping element zero, implying a virtual disconnection
of the side phenyl rings from the backbone. In our previous works\cite{shukla1,shukla2,shukla3},
we argued that the steric hindrance effects can be taken into account
by assuming that the phenyl rings of the unit cell are rotated with
respect to the $y$-axis by 30 degrees in such a manner that the oligomers
still have inversion symmetry. It is obvious that along the direction
of conjugation (longitudinal direction), PDPA is structurally similar
to \emph{trans-}polyacetylene, with alternating single and double
bonds. But, perpendicular to the conjugation direction (transverse
direction), the material has features in common with \emph{trans}-stilbene,
with its two phenyl rings, and vinylene-like linkage connecting them.
In the following, we will adopt the notation PDPA-$n$ to denote a
PDPA oligomer containing $n$ unit cells of the type depicted in Fig.
\ref{fig-pdpa}.

The point group symmetry associated with \emph{trans}-polyacetylene
(and polyenes) is $C_{2h}$ so that the one-photon states belong to
the irreducible representation (irrep) $B_{u}$, while the ground
state and the two-photon excited states belong to the irrep $A_{g}$.
Because of the phenyl group rotation mentioned above, the point group
symmetry of PDPA's is $C_{i}$ so that its ground state and the two-photon
excited states belong to the irrep $A_{g}$, while the one-photon
excited states belong to the irrep $A_{u}$. However, to facilitate
direct comparison with polyenes, we will refer to the one-/two-photon
states of PDPA's also as $B_{u}$/$A_{g}$-type states.

The independent-electron calculations on the oligomers PDPA-$n$ were
performed using the H\"{u}ckel model Hamiltonian which, adopting
a notation identical to our previous works\cite{shukla1,shukla2,shukla3}
reads, \begin{subequations}
\label{allequations}
\begin{equation}
H = H_C + H_P + H_{CP},  \label{eq-ham0}
\end{equation}
where \(H_C\) and \(H_P\) are the one-electron Hamiltonians for the carbon atoms 
located on the {\emph{trans}}-polyacetylene backbone (chain), and the  phenyl groups, respectively, 
H\(_{CP}\) is the one-electron hopping between the chain and the phenyl
units. The individual terms can now be written as, 
\begin{equation}
H_C = -\sum_{\langle k,k' \rangle,M} (t_0 - (-1)^M \Delta t) 
B_{k,k';M,M+1},
  \label{eq-h1}
\end{equation}
\begin{equation}
H_P=-t_0 \sum_{\langle\mu,\nu\rangle,M} B_{\mu,\nu;M,M}, \label{eq-h2}
\end{equation}
and 

\begin{equation}
H_{CP}= -t_{\perp} \sum_{\langle k,\mu \rangle,M} B_{k,\mu;M,M,}. \label{eq-h3}
\end{equation}
\end{subequations} In the equation above, $k$, $k'$ are carbon atoms on the polyene
backbone, $\mu,\nu$ are carbon atoms located on the phenyl groups,
$M$ is a unit consisting of a phenyl group and a polyene carbon,
$\langle...\rangle$ implies nearest neighbors, and $B_{i,j;M,M'}=\sum_{\sigma}(c_{i,M,\sigma}^{\dagger}c_{j,M',\sigma}+h.c.)$.
Matrix elements $t_{0}$, and $t_{\perp}$ depict one-electron hops.
In $H_{C}$, $\Delta t$ is the bond alternation parameter arising
due to electron-phonon coupling. In $H_{CP}$, the sum over $\mu$
is restricted to atoms of the phenyl groups that are directly bonded
to backbone carbon atoms. There is a strong possibility that due to
the closeness of the phenyl rings in the adjacent unit cells, there
will be nonzero hopping between them, giving rise to a term $H_{PP}$
in the Hamiltonian above. However, in our earlier study,\cite{shukla2}
we explored the influence of this coupling on the linear optics of
these materials, and found it to have insignificant influence. Therefore,
in the present study, we believe that we are justified in ignoring
the phenyl-phenyl coupling.

As far as the values of the hopping matrix elements are concerned,
we took $t_{0}=-2.4$ eV, while it is imperative to take a smaller
value for $t_{\perp}$, because of the twist in the corresponding
bond owing to the steric hindrance mentioned above. We concluded that
for a phenyl group rotation of 30 degrees, the maximum possible value
of $t_{\perp}$ can be -1.4 eV.~\cite{shukla1} Bond alternation
parameter $\Delta t=0.45$ eV was chosen so that the backbone corresponds
to \emph{trans}-polyacetylene with the optical gap of 1.8 eV in the
long chain limit.

The THG process in oligo-PDPA's were studied by computing the third-order
nonlinear susceptibilities $\chi^{(3)}(-3\omega;\omega,\omega,\omega)$.
For short we will refer to this susceptibility, as $\chi_{THG}^{(3)}$.
First the H\"{u}ckel Hamiltonian for the corresponding oligo-PDPA
was diagonalized to compute the one-electron eigenvalues and eigenfunctions.
These quantities were subsequently used in the formulas derived by
Yu and Su,~\cite{yu-su} to compute $\chi_{THG}^{(3)}$.~

The correlated calculations on the oligomers PDPA-$n$ were performed
using the P-P-P model Hamiltonian \begin{equation}
H=H_{C}+H_{P}+H_{CP}+H_{ee},\label{eq-ham}\end{equation}
 where $H_{C}$, $H_{P}$, $H_{CP}$ have been explained above. H$_{ee}$
depicts the electron-electron repulsion and can be written as \begin{eqnarray}
H_{ee} & = & U\sum_{i}n_{i\uparrow}n_{i\downarrow}\nonumber \\
 &  & +\frac{1}{2}\sum_{i\neq j}V_{i,j}(n_{i}-1)(n_{j}-1)\label{eq-hee}\end{eqnarray}
 where $i$ and $j$ represent all the atoms of the oligomer. The
Coulomb interactions are parameterized according to the Ohno relationship
\cite{ohno}, \begin{equation}
V_{i,j}=U/\kappa_{i,j}(1+0.6117R_{i,j}^{2})^{1/2}\;\mbox{,}\label{eq-ohno}\end{equation}

where, $\kappa_{i,j,M,N}$ depicts the dielectric constant of the
system which can simulate the effects of screening, $U$ is the on-site
repulsion term, and $R_{i,j}$ is the distance in \AA ~ between
the $i$th carbon and the the $j$th carbon. The Ohno parameterization
initially was carried out for small molecules, and, therefore, it
is possible that the Coulomb parameters for the polymeric samples
could be somewhat smaller due to interchain screening effects.\cite{chandross}
Since the results obtained will clearly depend on the choice of the
Coulomb parameters, we tried two sets: (a) {}``standard parameters''
with $U=11.13$ eV and $\kappa_{i,j}=1.0$, and (b) {}``screened
parameters'' with $U=8.0$ eV and $\kappa_{i,i,M,M}=1.0$, and $\kappa_{i,j,M,N}=2.0$,
otherwise. Using the screened parameters, Chandross and Mazumdar~\cite{chandross}
obtained better agreement with experiments on excitation energies
of PPV oligomers, as compared to the standard parameters. Recently,
we performed a large-scale correlated study of singlet and triplet
excited states in oligo-PPV's and observed a similar trend.~\cite{shukla-ppv}

As far as the hopping matrix elements for PDPA's and polyenes are
concerned, they were assigned the values used in the H\"uckel-model
calculations, except for the bond-alternation parameter for which
the smaller value $\Delta t=0.168$ eV, consistent with the P-P-P
model, was used. In \emph{trans}-stilbene, for phenyl rings hopping
value of -2.4 eV was used, while in the vinylene linkage the values
used were -2.2 eV for the single bond, and -2.6 eV for the double
bond. In all the calculations, C-C bond length of 1.4 \AA ~ was
used for the phenyl rings. In polyenes and PDPA's, along the backbone
the single bonds and the double bonds were taken to be 1.45 \AA~
and 1.35 \AA, respectively. The bond connecting the backbone to the
substituent phenyl rings was taken to be 1.40 \AA. In \emph{trans}-stilbene,
in the vinylene linkage, the single (double) bond lengths were taken
to be 1.54 \AA~ (1.33 \AA). The bond lengths used in this paper
are the same as the ones used in our earlier works.\cite{shukla1,shukla2,shukla3,shukla-ppv}

The starting point of the correlated calculations for various oligomers
were the restricted Hartree-Fock (HF) calculations, using the P-P-P
Hamiltonian. The many-body effects beyond HF were computed using different
levels of the configuration interaction (CI) method, namely, quadruples-CI
(QCI), and the multi-reference singles-doubles CI (MRSDCI). Since
the number of electrons in oligo-PDPA's is quite large despite the
P-P-P approximation owing to the large unit cell, it is not possible
to include all the orbitals in the many-body calculations. Therefore,
one has to reduce the number of degrees of freedom by removing some
orbitals from the many-body calculations. In order to achieve that,
for each oligomer we first decided as to which occupied and the virtual
orbitals will be active in the many-body calculations based upon:
(a) their single-particle HF energies with respect to the location
of the Fermi level, and (b) Mulliken populations of various orbitals
with respect to the chain/phenylene-based atoms. Because of the particle-hole
symmetry in the problem, the numbers of active occupied and virtual
orbitals were taken to be identical to each other, with the occupied
and virtual orbitals being particle-hole symmetric. The remaining
occupied orbitals were removed from the many-body calculations by
the act of {}``freezing'', i.e., by summing up their interactions
with the active electrons, and adding this effective potential to
the one-electron part of the total Hamiltonian. The inactive virtual
orbitals were simply deleted from the list of orbitals. When we present
the CI results on various oligo-PDPA's, we will also identify the
list of active orbitals. During the CI calculations, full use of the
spin and the point group ($C_{i}$ for PDPA's) symmetries was made.
From the CI calculations, we obtain the eigenfunctions and eigenvalues
corresponding to the correlated ground and excited states of various
oligomers. Using the many-body wave functions, we compute the matrix
elements of the dipole operator amongst various states. Finally, these
quantities are fed in to the sum-over-states formulas of Orr and Ward,~\cite{orr}
to obtain the correlated values of $\chi_{THG}^{(3)}$. More details
about the procedural aspects of various CI approaches used by us can
be found in our earlier works.~\cite{shukla2,shukla3,shukla-ppv,hng-ppv}

\section{Results and Discussion}

\label{results} In this section we present our results on $\chi_{THG}^{(3)}$
using various approaches. First we present and briefly discuss the
results computed using the H\"uckel model, followed by the results
computed using various CI approaches.

\subsection{Independent-electron theory}

\label{res-huck}

\subsubsection{Longitudinal Component}

\label{res-huck-x} Here we present and discuss the longitudinal THG
spectrum for the case of oligomer PDPA-50, which has 700 sites leading
to 350 occupied and the same number of virtual orbitals. Even within
the one-electron theory, calculation of $\chi^{(3)}$'s with such
a large number of orbitals is intractable. Therefore, to perform such
calculations on larger oligomers, it is mandatory that some of the
orbitals be discarded. To this end we adopted the criterion to discard
those occupied and virtual orbitals which are more than 3.0 eV away
from the Fermi energy. The choice of 3.0 eV is based upon the magnitude
of the optical gap and the location of various features in the $\chi^{(3)}$
spectra of smaller oligomers. Since, on PDPA-30 it was also possible
to perform exact calculations, we were able to compare our approximate
results with the exact ones for that oligomer. We found that the approximate
results on longer oligomers are quite accurate. 

\begin{figure}
\begin{center}\includegraphics[%
  width=8cm,
  angle=-90]{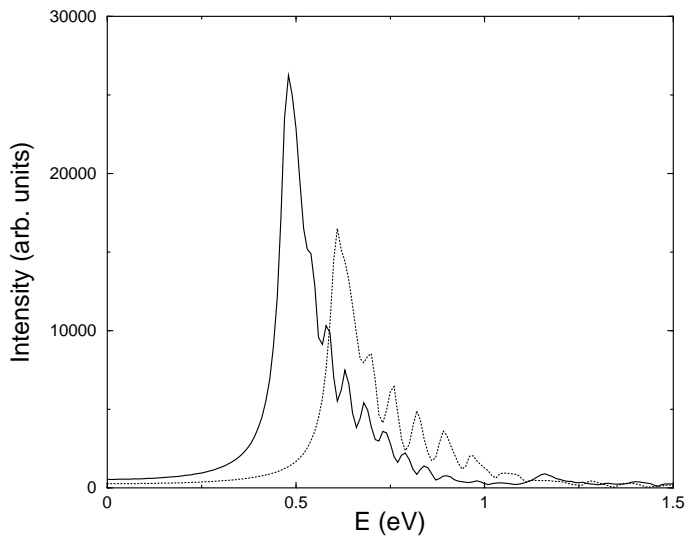}\end{center}

\caption{Comparison of $|\chi_{xxxx}^{(3)}(-3\omega;\omega,\omega,\omega)|$
of PDPA (solid line) and \emph{trans}-polyacetylene (dotted line)
oligomers containing fifty unit cells, computed using the H\"{u}ckel
model. Linewidth of 0.05 eV was assumed for all energy levels.}

\label{thg-pdpa-tpa}
\end{figure}

The total magnitude of component of the longitudinal component of
THG susceptibility ($|\chi_{xxxx}^{(3)}(-3\omega;\omega,\omega,\omega)|$),
as a function of the incident photon energy $E$, for fifty unit cell
oligomers of PDPA and \emph{trans}-polyacetylene is plotted in Fig.
\ref{thg-pdpa-tpa}. From the figure it is clear that the main peak
of the spectrum for both these materials is the first one corresponding
to the lowest optically allowed state $1B_{u}$, and is located at
$\hbar\omega=E_{g}/3$, where $E_{g}$ is the value of the optical
gap defined as $E_{g}=E(1B_{u})-E(1A_{g})$. From the figure it is
obvious that: (a) qualitatively speaking, longitudinal THG spectra
for the both the polymers are similar (b) from a quantitative point
of view, as expected, the resonant features of the PDPA spectrum are
red shifted as compared to the \emph{trans}-polyacetylene ones (c)
in the frequency region of interest, both the resonant, and the off-resonant,
nonlinear optical responses of PDPA are much more intense as compared
to \emph{trans}-polyacetylene. Since the $1B_{u}$ state of PDPA's
in H\"uckel model originates from the excitation $HOMO\rightarrow LUMO$,
both of which are chain-based orbitals, one can safely conclude that
as far as the longitudinal nonlinear-optical response computed at
the H\"uckel level is concerned, oligo-PDPA's behave as if they were
smaller band-gap polyenes.

\subsubsection{Transverse Component}

\label{res-huck-y} 
\begin{figure}
\begin{center}\includegraphics[%
  width=8cm,
  angle=-90]{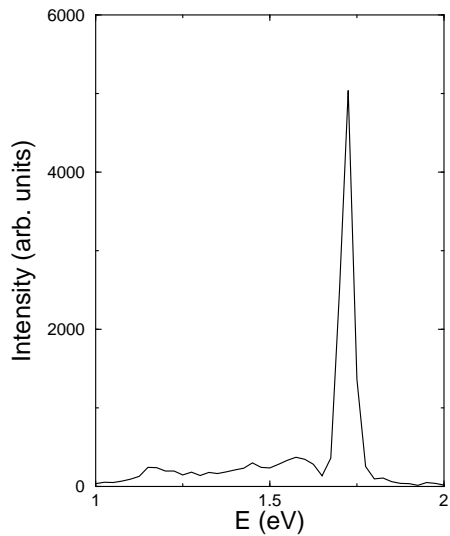}\end{center}

\caption{$|\chi_{yyyy}^{(3)}(-3\omega;\omega,\omega,\omega)|$ of PDPA-30
 plotted as a function of incident photon energy. Linewidth of 0.05
eV was assumed for all energy levels. }

\label{thg-pdpa30-y}
\end{figure}

The transverse THG spectrum of oligo-PDPA's saturates
very rapidly with size because with increasing conjugation length,
the \emph{trans}-stilbene-like structure of the oligomer in the transverse
direction is unchanged. With increasing conjugation length, only the
overall intensity in the transverse direction is expected to increase,
with little change in the qualitative feature. Therefore, here we
present the exact results on $|\chi_{yyyy}^{(3)}(-3\omega;\omega,\omega,\omega)|$
for the oligomer PDPA-30 in Fig. \ref{thg-pdpa30-y}. It is clear
from the figure that there is only one resonant feature in the spectrum
and it is located close to 1.7 eV and its intensity is about one-fifth
that of the largest resonant intensity in the longitudinal THG spectrum(cf.
Fig. \ref{thg-pdpa-tpa}). Even for smaller oligomers such as PDPA-10
and PDPA-20, the location of the resonance is the same pointing to
its possible origins in the phenyl-based orbitals. Therefore, to simplify
our task of locating these orbitals, we investigate the THG spectrum
of PDPA-10 for which the main peak occurs at 1.71 eV, corresponding
to an $A_{g}$ type state, obtained from the ground state by single-excitations
$H\rightarrow L+39$ and $H-39\rightarrow L$. To investigate the
nature of the orbital corresponding to the $L+39$-th one-electron
state, we calculated the contribution of the charge density centered
on the backbone carbon atoms, to its total normalization. The contribution
was computed to be $0.05$ which indicates that the orbital in question
is predominantly centered on the side phenyl rings. Further investigation
of the orbital coefficient reveals that the orbitals in question,
are derived from the phenyl-based delocalized ($d^{*})$ virtual orbitals,
with no contribution from the localized orbitals ($l^{*}$-type) of
the phenyl rings. Similarly, owing to the particle-hole symmetry,
$H-39$-th orbital is derived from the $d$-type occupied orbitals
of the phenyl rings. The investigation of longer oligomers yields
identical results. Thus it is clear that the transverse THG susceptibility
of the system owes its origins to phenyl-based delocalized ($d/d^{*}$)
levels. This is an important point which also helps us perform effective
correlated calculations of this component, presented in the next section.

\subsection{Correlated-electron theory}

\label{res-ci}Since the smaller energy gaps obtained with the screened
parameters in our earlier works were found to be in much better agreement
with the experiments, we will present our main results based upon
screened-parameter-based calculations. However, when we compare the
PDPA nonlinear optical spectra with those of polyenes, we will use
the standard parameters because screened parameters are not valid
for polyenes.

\subsubsection{Longitudinal Component}

\label{res-thg-x} Now we present correlated calculations for the
longitudinal component of the third-harmonic generation susceptibility
$\chi_{xxxx}^{(3)}(-3\omega;\omega,\omega,\omega)$ for oligo-PDPA's,
performed using the QCI method. Given the large number of electrons
in these systems, QCI method is not feasible for them if all the orbitals
of the system are retained in the calculations. Since the longitudinal
nonlinear optical properties are determined by low-lying excited states
of the system, in the limited CI calculations we decided to include
the orbitals closest to the Fermi level. Therefore, for PDPA-$n$
we included $n$ occupied, and $n$ virtual orbitals closest to the
Fermi level in the QCI calculations. Remaining occupied orbitals were
frozen and virtual orbitals were deleted as explained in section \ref{method}.
Thus, the computational effort associated with the QCI calculations
on PDPA-$n$ is same as that needed for a polyene with $n$ double
bonds. Although for PDPA-$10$, it leads to Hilbert space dimensions
in excess of one million, however, using the methodology reported
in our earlier works\cite{shukla-ppv,shukla2}, we were able to obtain
low-lying excited states of such systems.

\begin{figure}
\begin{center}\includegraphics[%
  width=12cm,
  angle=-90]{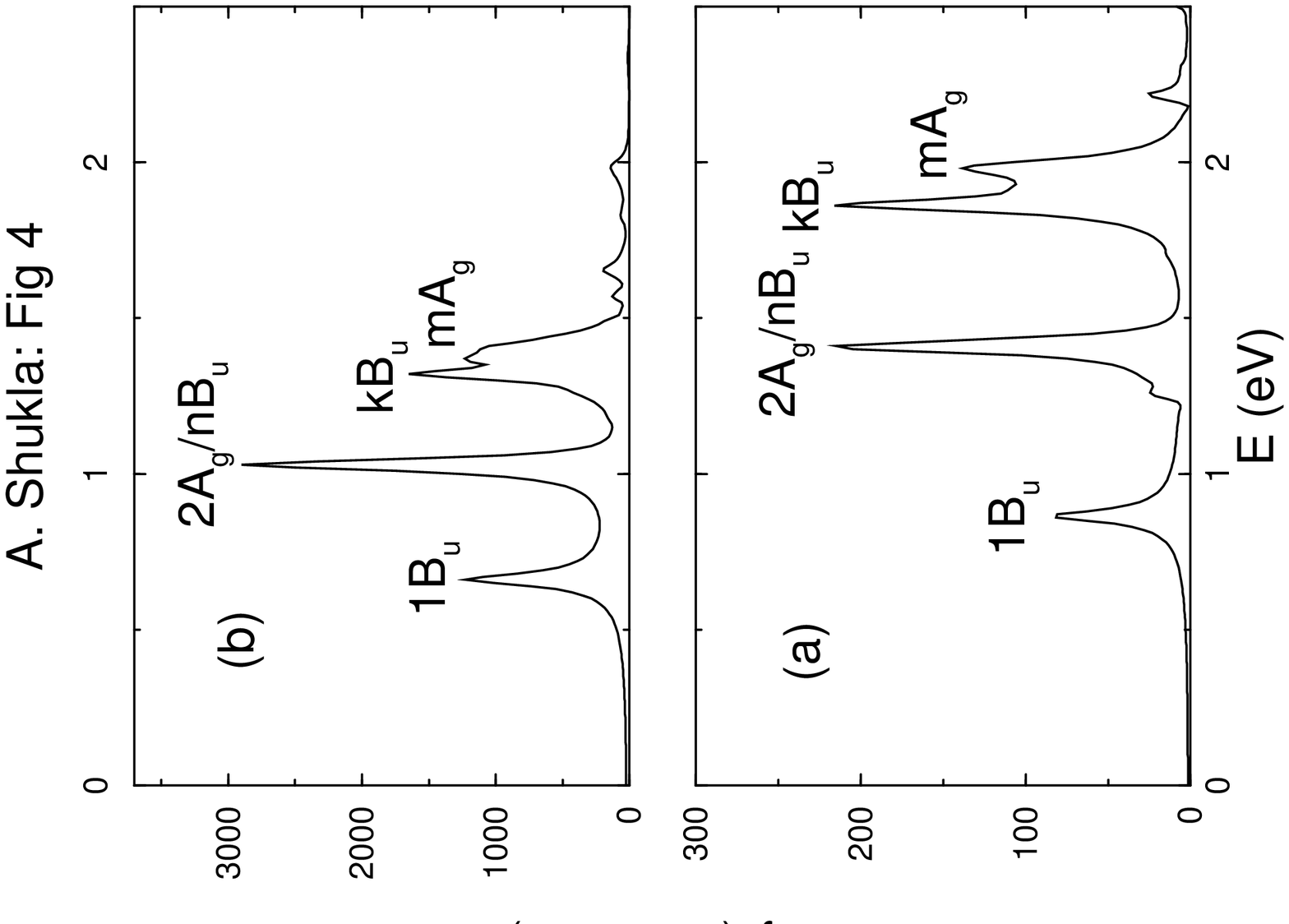}\end{center}

\caption{The magnitudes of the longitudinal component of the third-harmonic
generation spectra ($|\chi_{xxxx}^{(3)}(-3\omega;\omega,\omega,\omega)|$)
of oligo-PDPA's computed using the QCI approach and the screened parameters
for: (a) PDPA-5 (b) PDPA-10. A linewidth of 0.05 eV was assumed for
all the levels.}

\label{pdpa-qci-thg}
\end{figure}

The longitudinal THG spectra ($|\chi_{xxxx}^{(3)}(-3\omega;\omega,\omega,\omega)|$)
for PDPA-5 and PDPA-10 employing the screened parameters and the QCI
method are presented in Figs.\ref{pdpa-qci-thg}(a) and \ref{pdpa-qci-thg}(b),
respectively. From the figures it is obvious that the spectra for
PDPA-5 and PDPA-10 containing four main features each are qualitatively
quite similar, suggesting the possibility that an oligomer as small
as PDPA-5 may possess essential features of bulk PDPA. It is clear
from Fig. \ref{pdpa-qci-thg} that: (a) The intensity of the nonlinear
response increases with the increasing length of the oligomer, and
(b) the peaks of the oligomers are redshifted with the increasing
conjugation length. Next we examine these peaks in detail.

\begin{figure}
\begin{center}\includegraphics[%
  width=12cm,
  angle=-90]{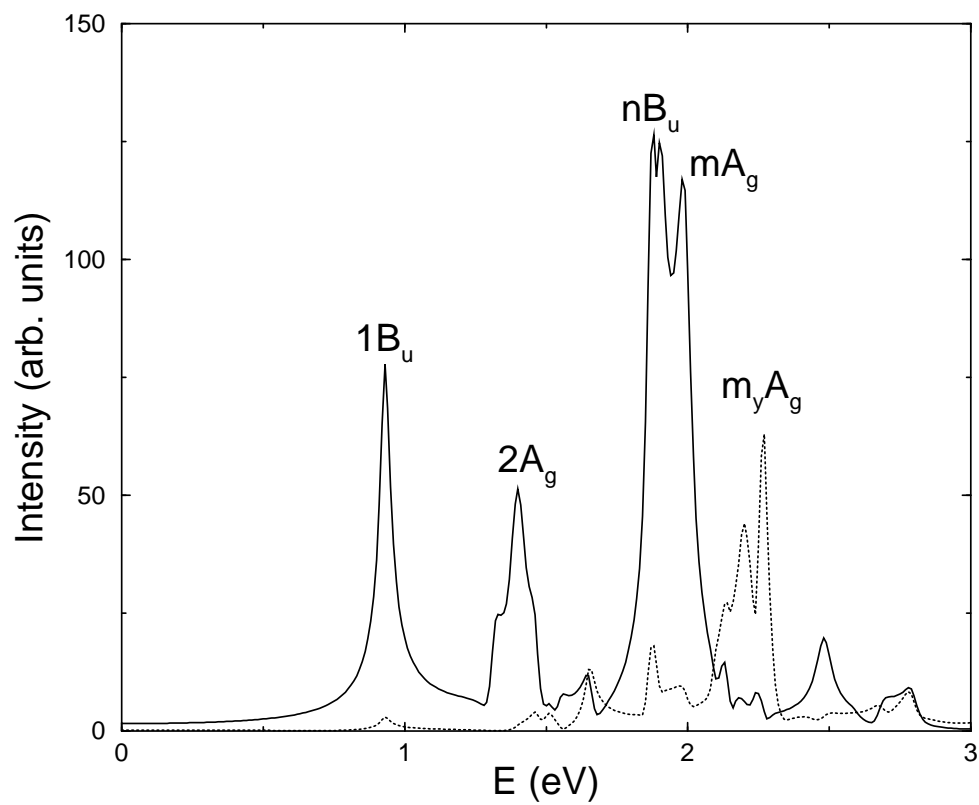}\end{center}

\caption{Comparison of longitudinal (solid lines ) and the transverse (dotted
lines) third-harmonic generation spectra of PDPA-5 computed using
the screened parameters and the MRSDCI method. A linewidth of 0.05
eV was assumed for all the levels. }

\label{pdpa_thg_x_y}
\end{figure}

The properties of excited states contributing to the longitudinal
THG spectra of PDPA-5 and PDPA-10 are presented in table \ref{tab-pdpa5}.
From the table, and Fig. \ref{pdpa-qci-thg}, it is clear that there
are three $B_{u}$-type states, $1B_{u}$, $nB_{u}$, and $kB_{u}$
along with two $A_{g}$-type states, $2A_{g}$ and $mA_{g}$ which
contribute, respectively, to the three-photon and two-photon resonances
in the spectra. Upon comparing the wave functions of various excited
states of PDPA-5, with those of PDPA-10, we infer that qualitatively
the wave functions are very similar. For each oligomer, $1B_{u}$
state, which constitutes the first peak of the spectrum consists mainly
of across-the-gap, single excitation. The next peak corresponds to
the $nB_{u}$ state which, for both the oligomers, also consists mainly
of higher energy single excitations. The third three-photon resonance
corresponding to a still higher energy state labeled $kB_{u}$, is
a mixture of both singly- and doubly-excited configurations. For PDPA-10,
this state is dominated by a high-energy single excitation with important
contributions from double excitations, while in PDPA-5 the situation
is reverse with main contributions coming from the double excitations.
Next, when we examine the wave functions of the states $2A_{g}$ and
$mA_{g}$ contributing to the two-photon features, we find that configurations
contributing to them are essentially identical. However, because of
the orthogonality constraint, the relative signs of the coefficients
of these configurations are opposite for the two states. One strange
aspect of the THG spectrum presented in Fig. \ref{pdpa-qci-thg} is
the coincidence of the two-photon and three-photon resonances corresponding
to states $nB_{u}$ and $2A_{g}$, respectively, for both PDPA-5 and
PDPA-10. This is because for both the oligomers, $E(nB_{u})/3\approx E(2A_{g})/2$.
One wonders whether this is an artifact of using a truncated orbital
set in the QCI calculations, or a genuine effect. In order to perform
correlated calculations of the transverse THG spectrum of PDPA-5,
we used an extended orbital set and performed large-scale MRSDCI calculations
presented later in Fig. \ref{pdpa_thg_x_y}. From there it is clear
that although both $2A_{g}$ and $nB_{u}$ make important contributions
to the spectrum, however, their peaks are well separated, and the
intensity of $nB_{u}$ is much higher than that of $2A_{g}$, leading
us to conclude that the coincidence of $2A_{g}/nB_{u}$ peaks in the
QCI/THG spectrum is an artifact.

\begin{figure}
\begin{center}\includegraphics[%
  width=12cm,
  angle=-90]{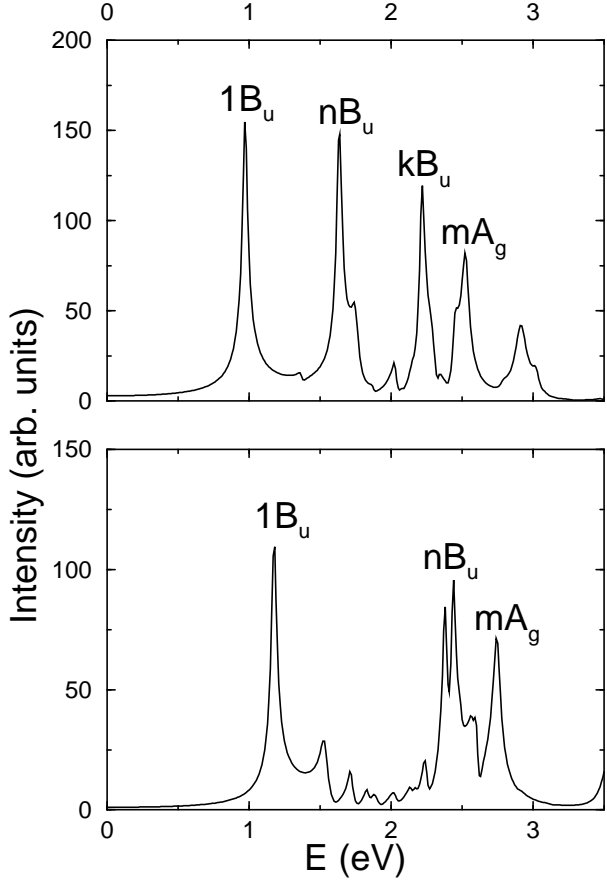} \end{center}

\caption{The magnitude of the third-harmonic generation spectrum ($|\chi_{xxxx}^{(3)}(-3\omega;\omega,\omega,\omega)|$)
of: (a) PDPA-10 (top) and (b) ten unit oligomer of \emph{trans}-polyacetylene
(bottom) computed using the QCI approach and the standard parameters.
A linewidth of 0.05 eV was assumed for all the levels.}

\label{thg-tpa-pdpa}
\end{figure}

Next we compare the longitudinal THG spectrum of PDPA-10 with that
of a polyene consisting of ten double bonds (PA-10), computed using
the QCI method, and the standard parameters in the P-P-P Hamiltonian.
The spectra are presented in Fig. \ref{thg-tpa-pdpa}, while the properties
of the excited states contributing to the spectra are illustrated
in table \ref{tab-pdpa-tpa}. It is clear from that the intensity
of the PDPA-10 spectrum is generally larger and all its major peaks
are significantly redshifted as compared to the \emph{trans}-polyacetylene
oligomer, in agreement with the H\"uckel model results. Moreover,
the THG spectrum of PDPA-10 computed with the standard parameters,
still has four main peak corresponding to the states $1B_{u}$, $nB_{u}$,
$kB_{u}$, and $mA_{g}$, in agreement with the screened-parameter
results. The main difference in the THG spectrum of PDPA-10 computed
with standard parameters, as compared to the one computed with the
screened parameters, is that there is no contribution to it from the
$2A_{g}$ state. When we compare this aspect of PDPA-10 standard parameter
THG spectrum with that of PA-10, we find that the PA-10 spectrum also
does not exhibit any feature corresponding to $2A_{g}$. Moreover,
PA-10 THG spectrum has only three peaks corresponding to $1B_{u}$,
$nB_{u}$, and $mA_{g}$, with no peak corresponding to a $kB_{u}$-type
state. If we compare the wave functions of these three states of PA-10
with those of PDPA-10 (table \ref{tab-pdpa-tpa}), we find that the
wave functions of $1B_{u}$ and $mA_{g}$ states are very similar
to each other for the two materials. As far as the $nB_{u}$ state
is concerned, its wave function for PDPA-10 is composed predominantly
of single excitations, while for PA-10 it exhibits large configuration
mixing, with significant contribution also coming from the double
excitations. Moreover, in PDPA-10, the $nB_{u}$ state is very close
energetically to the $mA_{g}$ state (slightly below it), while in
PA-10, it is much higher in energy than the $nB_{u}$ state. Note
that the $nB_{u}$ state of PDPA-10 computed with the screened parameters
is slightly above the $mA_{g}$ state (table \ref{tab-pdpa5}). Therefore,
the location of the $nB_{u}$ state is an interesting difference between
PA-10 and PDPA-10 which these standard-parameter-based calculations
suggest, and it merits further investigation in possibly more extensive
calculations. Thus, experimental investigations of the longitudinal
THG spectrum of PDPA-10 could possibly shed some light on the following
aspects: (a) whether or not the $2A_{g}$ state makes an important
contribution to the spectrum indicating as to what range of Coulomb
parameters (screened/standard) are valid for the material, (b) location
of $nB_{u}$ vis-a-vis $mA_{g}$, and (c) whether or not there is
a peak corresponding to the $kB_{u}$ state in PDPA's.

\subsubsection{Transverse Component}

\label{res-thg-y} Performing accurate correlated calculations of
$\chi_{yyyy}^{(3)}(-3\omega;\omega,\omega,\omega)$ for oligo-PDPA's
is an extremely difficult task. The reason being that the many-body
$A_{g}$-type states which contribute to the peaks in this component
of the susceptibility are very high in energy, and hence are more
difficult to compute by many-body methods, as compared to the low-lying
excited states contributing to the longitudinal spectra. The very
high excitation energies of these states are due to the fact that
these $A_{g}$-type states are predominantly composed of excited configurations
involving high-energy delocalized orbitals originating from the phenyl
substituents. However, during our independent-electron study, we concluded
that transverse susceptibilities exhibit rapid saturation with the
conjugation length. Therefore, we restrict our study of the transverse
components of the nonlinear susceptibilities to PDPA-5. We perform
calculations with the screened parameters and the MRSDCI method employing
thirty orbitals in all, of which fifteen were occupied orbitals, and
the remaining fifteen were the virtual ones. Rest of the occupied
orbitals were frozen. Of the thirty orbitals, ten were the orbitals
closest to the Fermi level which were also used in the QCI calculations.
Remaining twenty orbitals were the $d/d^{*}$-type phenylene-ring-based
orbitals closest to the Fermi level. In the MRSDCI calculations, no
$l/l^{*}$-type phenylene orbitals were used because they do not make
any significant contribution to the transverse nonlinear spectra in
the H\"{u}ckel model calculations. In the MRSDCI calculations, we
used 25 reference configurations for the $A_{g}$-type states, and
24 for the $B_{u}$-type states leading to CI matrices of dimensions
close to half-a-million both for the $A_{g}$ and $B_{u}$ manifolds.

\begin{figure}
\begin{center}\includegraphics[%
  width=12cm,
  angle=-90]{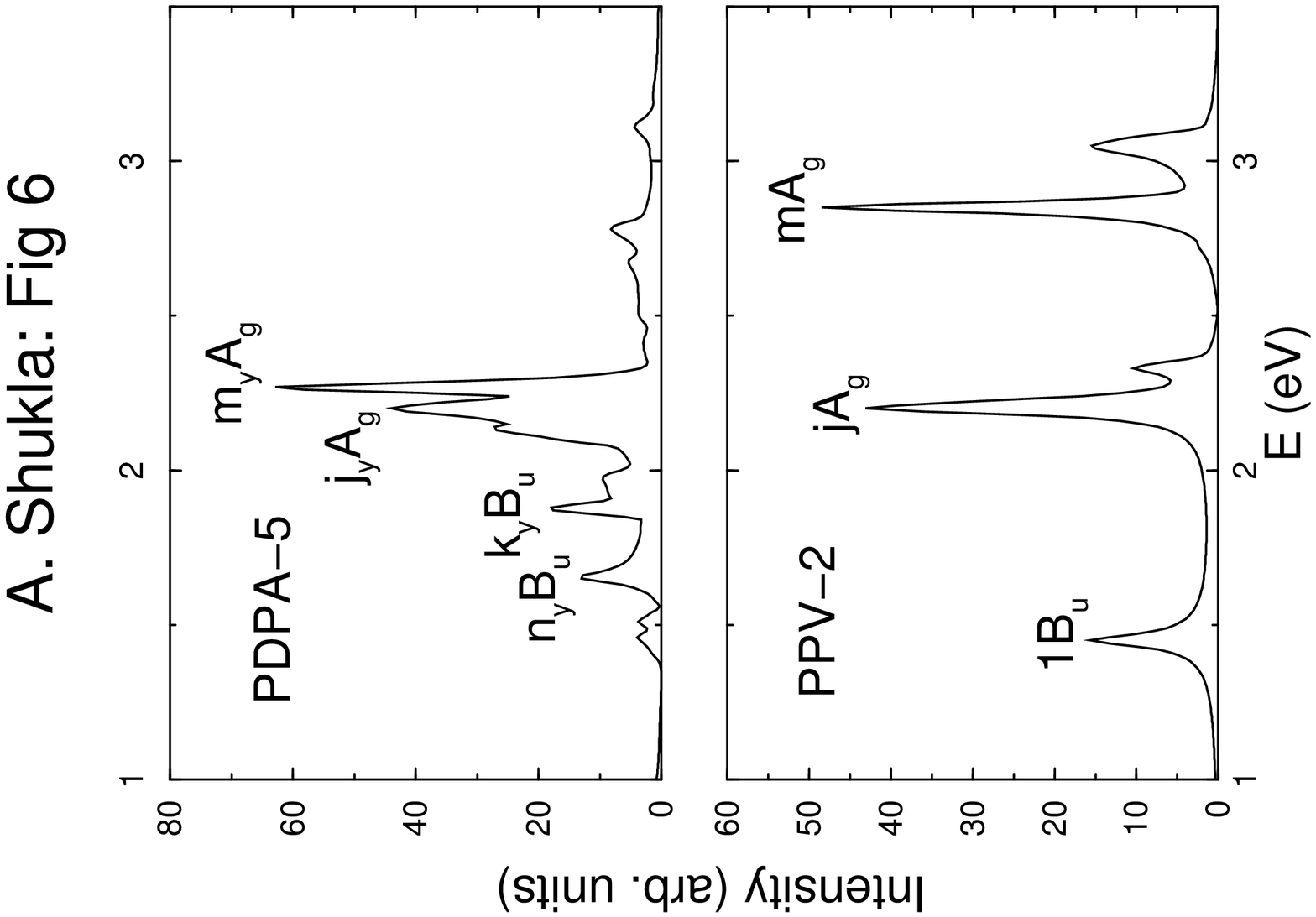}\end{center}

\caption{Comparison of  transverse  third-harmonic generation spectra of PDPA-5
with that of \emph{trans}-stilbene (PPV-2) oriented along the y-axis.
Both the spectra were computed using the screened parameters. For
PDPA-5, MRSDCI method was used, while for PPV-2, QCI method was used.
A linewidth of 0.05 eV was assumed for all the levels. }

\label{pdpa_ppv2_y}
\end{figure}

The transverse THG spectrum of PDPA-5, along with the THG spectrum
of \emph{trans}-stilbene (labeled PPV-2) oriented along the $y$-axis
computed using the QCI approach and screened parameters, are presented
in Fig. \ref{pdpa_ppv2_y}. The properties of the excited states contributing
to these spectra are displayed in table \ref{tab-pdpa5-y}. Before
discussing the many-particle wave functions of various excited states,
it is important to specify that for PDPA-5, orbitals $H-17/L+17$
and $H-19/L+19$ appearing in various states are the substituent phenyl-ring-based
$d/d^{*}$ orbitals making important contributions to the transverse
THG spectra computed using the H\"uckel model (cf. sec. \ref{res-huck-y}).
It is obvious from Fig. \ref{pdpa_ppv2_y} that: (a) For PDPA-5 four
states labeled $n_{y}B_{u}$, $k_{y}B_{u}$, $j_{y}A_{g}$, and $m_{y}A_{g}$
(subscript $y$ implies $y,$ i.e., transverse direction) contribute
to the spectrum, while (b) for PPV-2 three states $1B_{u}$, $jA_{g}$,
and $mA_{g}$ make main contributions. Furthermore, the main contributions
to the intensities of the THG spectra of both the substances are from
two $A_{g}$-type peaks: $j_{y}A_{g}$ and $m_{y}A_{g}$ for PDPA-5,
and $jA_{g}$ and $mA_{g}$ for PPV-2. For PDPA, we recall that at
the H\"uckel model level, the transverse THG spectrum had only one
peak due to closely-spaced $A_{g}$-type states corresponding to a
single-particle excitation from the orbitals near the Fermi level
($H/L)$ to the phenyl-based $d/d^{*}$ orbitals. Upon examining the
many-particle wave functions of various excited states of PDPA-5,
we conclude that even at the many-body level the excited states contributing
to the transverse THG spectra originate from the same set of single
excitations involving $H/L$ and phenyl-based $d/d^{*}$ orbitals
except for the $k_{y}B_{u}$state. However, the charge-density analysis
of the orbitals involved in the $k_{y}B_{u}$ state reveals that the
orbitals in question ($H-3/L+3,\: H-4/L+4$) have significant charge
density on the substituent phenyls. Thus, we can safely conclude that
in the transverse THG spectrum of PDPA-5, the substituent-phenyl-based
orbitals play a very important role. On comparing the many-particle
wave functions of the $A_{g}$-type states contributing to the THG
spectra of the two systems, we find that unlike the case of PDPA-5,
in PPV-2 these states ($kA_{g}$ and $mA_{g}$) have important contributions
from the two-particle excitations as well. However, the contributions
of the single-particle excitations to these states in both PDPA-5
and PPV-2 have similar character---they involve $H/L$ orbitals and
high-lying orbitals localized predominantly in the phenyl rings. Upon
further investigation of these high-lying orbitals involved in single
excitations we conclude that they are of $d/d^{*}$-type phenyl-based
orbitals. Therefore, we conclude that the mechanism of the transverse
THG in PDPA's has noticeable similarities to that of THG in \emph{trans}-stilbene. 

Finally, in Fig. \ref{pdpa_thg_x_y}, we compare the transverse THG
spectrum of PDPA-5 with its longitudinal one, both computed using
the MRSDCI procedure, and the screened parameters. It is clear from
the figure that the most intense peak in the transverse spectrum is
about half in intensity as compared to the strongest peak in the longitudinal
one, and is located in the higher energy range where there is no competing
longitudinal response. Thus, our PDPA-5 calculations suggest that
the transverse THG response of these materials, compared to their
longitudinal response, is significant, as well as clearly distinguishable
from it, in an experiment. However, it remains to be verified whether
the results obtained for small oligomers will also be valid for the
bulk PDPA's.

\section{Conclusions }

\label{conclusion}

Our aim behind undertaking the present theoretical study of the nonlinear
optical properties of the novel polymer PDPA was, not only to calculate
their THG spectra, but also to understand the underlying mechanism
in a way similar to what has been possible for simpler polymers such
as \emph{trans}-polyacetylene,\cite{dixit,sumit} PPP, and PPV.\cite{lavren,hng-ppv}
This understanding has chiefly arisen from the essential state mechanism
alluded to earlier in section \ref{intro}, which aims to explain
the nonlinear optical properties of chain-like,\cite{dixit,sumit}
and more recently phenylene-based conjugated polymers,\cite{lavren}
in terms of a small number of excited states. For PDPA's, whose structure
has ingredients in common with both the chain-like as well as the
phenyl-based polymers, our calculations suggest that their nonlinear
optical properties can also be understood along the same lines as
the essential state mechanism. However, because of the anisotropic
nonlinear optical response of PDPA's stemming from their structural
anisotropy, the essential states contributing to various components
of susceptibility are different. Our independent-electron as well
as correlated calculations suggest that the THG spectrum of PDPA's
is distributed over two natural energy scales: (a) low-energy region
of the spectrum has significant intensity mainly for the longitudinal
component, while (b) high-energy region of the spectrum has intensity
mostly for the transverse component. The longitudinal THG spectrum
bears striking qualitative resemblance to the THG spectrum of polyenes
with similar essential states contributing to the intensities therein.
The transverse THG spectrum on the other hand, has resemblance to
the THG spectrum of \emph{trans-}stilbene, with again, similar type
of essential states contributing to their intensities. The essential
states which contribute to the longitudinal spectra of these polymers
are quite distinct from the states contributing to their transverse
spectra. The states contributing to the longitudinal spectra can be
described in terms of chain-based orbitals close to the Fermi level,
however, to describe the transverse spectra, one needs to take into
account the excitations involving phenyl-based orbitals away from
the Fermi level. Therefore, it will be of considerable interest if
the theoretical predictions presented here can be tested in experiments.

\section{Acknowledgments}

We are grateful to Sumit Mazumdar (University of Arizona) for a critical
reading of the manuscript, and for many suggestions for improvement.
These calculations were performed on the Alpha workstations of Physics
Department, and the Computer Center, IIT Bombay.

\begin{table}

\caption{Excited states contributing to the longitudinal THG spectrum of PDPA-5
and PDPA-10 as obtained in QCI calculations. Under the heading wave
function, we list the most important configurations contributing to
the many-body wave function of the state concerned, along with their
coefficients, consistent with our convention\cite{notation}. }

\begin{tabular}{|c|l|l|l|l|}
\hline 
&
\multicolumn{2}{c|}{PDPA-5}&
\multicolumn{2}{c|}{PDPA-10}\tabularnewline
\hline
State&
Energy (eV)&
Wave Function&
Energy (eV)&
Wave Function\tabularnewline
\hline
\hline 
$2A_{g}$&
2.86&
$|H\rightarrow L+1\rangle+c.c.\:(0.57)$&
2.08&
$|H\rightarrow L+1\rangle+c.c.\:(0.49)$\tabularnewline
&
&
 $|H\rightarrow L;H\rightarrow L\rangle(0.48)$&
&
$|H\rightarrow L;H\rightarrow L\rangle(0.50)$\tabularnewline
\hline 
$mA_{g}$&
3.98&
$|H\rightarrow L+1\rangle+c.c.(0.37)$&
2.82 ($3A_{g})$&
$|H\rightarrow L+1\rangle+c.c.(0.68)$\tabularnewline
&
&
$|H\rightarrow L;H\rightarrow L\rangle(0.74)$&
2.88 ($4A_{g}$)&
$|H\rightarrow L;H\rightarrow L\rangle(0.37)$\tabularnewline
\hline
$1B_{u}$&
2.59&
$|H\rightarrow L\rangle(0.99)$&
1.98&
$|H\rightarrow L\rangle(0.96)$\tabularnewline
\hline 
$nB_{u}$&
4.21&
$|H-1\rightarrow L+1\rangle(0.94)$&
3.08&
$|H-1\rightarrow L+1\rangle(0.72)$\tabularnewline
&
&
$|H\rightarrow L;H\rightarrow L+1\rangle+c.c.(0.14)$&
&
$|H\rightarrow L+2\rangle+c.c.(0.38)$\tabularnewline
\hline 
$kB_{u}$&
5.58&
$|H\rightarrow L;H\rightarrow L+1\rangle+c.c.(0.57)$&
3.96&
$|H-2\rightarrow L+2\rangle(0.59)$\tabularnewline
&
&
$|H-1\rightarrow L+1;H-1\rightarrow L\rangle+c.c.(0.25)$&
&
$|H\rightarrow L;H\rightarrow L+1\rangle+c.c.(0.31)$\tabularnewline
&
&
$|H-1\rightarrow L+1\rangle(0.19)$&
&
$|H-1\rightarrow L+1;H-1\rightarrow L\rangle+c.c.(0.23)$\tabularnewline
\hline
\end{tabular}

\label{tab-pdpa5}
\end{table}

\begin{table}

\caption{Comparison of excited states of ten unit polyene (PA-10) and PDPA-10
computed using the QCI method and the standard parameters in the P-P-P
Hamiltonian. Rest of the information is the same as given in the caption
of table \ref{tab-pdpa5}.}

\begin{tabular}{|c|l|l|l|l|}
\hline 
&
\multicolumn{2}{c|}{PDPA-10}&
\multicolumn{2}{c|}{PA-10}\tabularnewline
\hline
\hline 
State&
Energy (eV)&
Wave Function&
Energy (eV)&
Wave Function\tabularnewline
\hline
\hline 
$1B_{u}$&
2.91&
$|H\rightarrow L\rangle(0.90)$&
3.53&
$|H\rightarrow L\rangle(0.82)$\tabularnewline
&
&
$|H-1\rightarrow L+1\rangle(0.32)$&
&
$|H-1\rightarrow L+1\rangle(0.28)$\tabularnewline
\hline 
$mA_{g}$&
5.05&
$|H\rightarrow L;H\rightarrow L\rangle(0.47)$&
5.49&
$|H\rightarrow L+1\rangle+c.c.(0.42)$\tabularnewline
&
&
$|H\rightarrow L;H-1\rightarrow L+1\rangle(0.45)$&
&
$|H\rightarrow L;H\rightarrow L\rangle(0.36)$\tabularnewline
&
&
$|H\rightarrow L+1\rangle+c.c.(0.33)$&
&
$|H\rightarrow L;H-1\rightarrow L+1\rangle(0.31)$\tabularnewline
\hline 
$nB_{u}$&
4.90&
$|H-1\rightarrow L+1\rangle(0.62)$&
7.31&
$|H\rightarrow L;H\rightarrow L+1\rangle+c.c.(0.28)$\tabularnewline
&
&
$|H\rightarrow L+2\rangle+c.c.(0.29)$&
&
$|H\rightarrow L;H-1\rightarrow L+2\rangle+c.c.(0.27)$\tabularnewline
&
&
&
&
$|H-1\rightarrow L+1\rangle(0.23)$\tabularnewline
\hline 
$kB_{u}$&
6.65&
$|H\rightarrow L;H\rightarrow L+1\rangle+c.c.(0.36)$&
&
\tabularnewline
&
&
$|H-1\rightarrow L+1\rangle(0.29)$&
&
\tabularnewline
&
&
$|H-2\rightarrow L+2\rangle(0.20)$&
&
\tabularnewline
\hline
\end{tabular}\label{tab-pdpa-tpa}
\end{table}
\begin{table}

\caption{Nature of excited states contributing to the transverse THG spectrum
of PDPA-5 computed by MRSDCI method. Rest of the information is same
as given in the caption of table \ref{tab-pdpa5}.}

\begin{tabular}{|c|l|l|c|l|l|}
\hline 
\multicolumn{3}{|c|}{PDPA-5}&
\multicolumn{3}{c|}{PPV-2}\tabularnewline
\hline
\hline 
State&
Energy (eV)&
Wave Function&
State&
Energy (eV)&
Wave Function\tabularnewline
\hline
\hline 
$j_{y}A_{g}$&
4.39&
$|H\rightarrow L+17\rangle+c.c\:(0.43)$&
$jA_{g}$&
4.45&
$|H\rightarrow L+3\rangle+c.c\:(0.43)$\tabularnewline
&
&
$|H\rightarrow L+19\rangle+c.c\:(0.43)$&
&
&
$|H\rightarrow L;H\rightarrow L\rangle+c.c\:(0.42)$\tabularnewline
\hline 
$m_{y}A_{g}$&
4.52&
$|H\rightarrow L+17\rangle+c.c\:(0.27)$&
$mA_{g}$&
5.72&
$|H-1\rightarrow L+1;H\rightarrow L\rangle(0.29)$\tabularnewline
&
&
$|H\rightarrow L+19\rangle+c.c\:(0.38)$&
&
&
$|H-2\rightarrow L+2;H\rightarrow L\rangle(0.29)$\tabularnewline
&
&
$|H-2\rightarrow L+1\rangle+c.c\:(0.30)$&
&
&
$|H\rightarrow L+3\rangle+c.c.(0.26)$\tabularnewline
&
&
&
&
&
$|H\rightarrow L+5\rangle+c.c.(0.25)$\tabularnewline
\hline
$n_{y}B_{u}$&
4.96&
$|H\rightarrow L+19\rangle+c.c.(0.60)$&
$1B_{u}$&
4.35&
$|H\rightarrow L\rangle(0.90)$\tabularnewline
\hline
$k_{y}B_{u}$&
5.61&
$|H-3\rightarrow L+3\rangle(0.56)$&
&
&
\tabularnewline
&
&
$|H-4\rightarrow L+4\rangle(0.49)$&
&
&
\tabularnewline
\hline
\end{tabular}

\label{tab-pdpa5-y}
\end{table}


\begin{thebibliography}{10}
\bibitem{prasad}See, e.g., P.N Prasad and D.J. Williams, \emph{Introduction to Nonlinear
Optical Effects in Molecules and Polymers} (Wiley, New York, 1991).
\bibitem{dixit}S.N. Dixit, D. Guo, and S. Mazumdar, Phys. Rev. B \textbf{43}, 6781
(1991). 
\bibitem{sumit}S. Mazumdar and F. Guo, J. Chem. Phys. \textbf{100}, 1665 (1994). 
\bibitem{shakin}V.A. Shakin, S. Abe, and T. Kobayashi, Phys. Rev. B \textbf{53}, 10656
(1996). 
\bibitem{tretiak}S. Tretiak, V. Chernyak, and S. Mukamel, Phys. Rev. Letts. \textbf{77},
4656 (1996). 
\bibitem{yaron}D. Yaron, Phys. Rev. B \textbf{54}, 4609 (1996). 
\bibitem{shuai}Z. Shuai, J.L. Bre'das, A. Saxena, and A.R. Bishop, J. Chem. Phys.
\textbf{109}, 2549 (1998). 
\bibitem{aparna}A. Chakrabarti and S. Mazumdar, Phys. Rev. B \textbf{59}, 4839 (1999). 
\bibitem{lavren}M. Y. Lavrentiev, W. Barford, S.J. Martin, H. Daly, and R.J. Bursill,
Phys. Rev. B \textbf{59}, 9987 (1999). 
\bibitem{tada1}K. Tada, R. Hidayat, M. Hirohata, M. Teraguchi, T. Masuda and K. Yoshino,
Jpn. J. Appl. Phys., part 2 \textbf{35}, L1138 (1996). 
\bibitem{tada2}K. Tada, R. Hidayat, M. Hirohata, H. Kajii, S. Tatsuhara, A. Fujii,
M. Ozaki, M. Teraguchi, T. Masuda and K. Yoshino, Proc. SPIE-Int.
Soc. Opt. Eng., \textbf{3145}, 171 (1997). 
\bibitem{liess}M. Liess, I. Gontia, T. Masuda, K. Yoshino, and Z.V. Vardeny, Proc.
SPIE-Int. Soc. Opt. Eng., \textbf{3145}, 179 (1997). 
\bibitem{fujii1}A. Fujii, M. Shkunov, Z.V. Vardeny, K. Tada, K. Yoshino, M. Teraguchi
and T. Masuda, Proc. SPIE-Int. Soc. Opt. Eng., \textbf{3145}, 533
(1997). 
\bibitem{gontia}I. Gontia, S.V. Frolov, M. Liess, E. Ehrenfreund, Z.V. Vardeny, K.
Tada, H. Kajii, R. Hidayat, A. Fujii, K. Yoshino, M. Teraguchi and
T.Masuda, Phys. Rev. Lett. \textbf{82}, 4058 (1999). 
\bibitem{sun}R. Sun, Y. Wang, X. Zou, M. Fahlam, Q. Zheng, T. Kobayashi, T. Masuda
and A.J. Epstein, Proc. SPIE-Int. Soc. Opt. Eng., \textbf{3476}, 332
(1998). 
\bibitem{hidayat}R. Hidayat, S. Tatsuhara, D.W. Kim, M. Ozaki, K. Yoshino, M. Teraguchi
and T. Masuda, Phys. Rev. B \textbf{61}, 10167 (2000). 
\bibitem{fujii2}A. Fujii, R. Hidayat, T. Sonoda, T. Fujisawa, M. Ozaki, Z.V. Vardeny,
M. Teraguchi, T. Masuda, and K. Yoshino, Synth. Met. \textbf{116},
95 (2001). 
\bibitem{shukla1}A. Shukla and S. Mazumdar, Phys. Rev. Lett \textbf{83}, 3944 (1999). 
\bibitem{shukla2}H. Ghosh, A. Shukla, and S. Mazumdar, Phys. Rev. B \textbf{62}, 12763
(2000). 
\bibitem{shukla3}A. Shukla, H. Ghosh, and S. Mazumdar, Synth. Met. \textbf{116}, 87
(2001). 
\bibitem{ohno}K. Ohno, Theor. Chim. Acta \textbf{2}, 219 (1964). 
\bibitem{chandross}M. Chandross and S. Mazumdar, Phys. Rev. B \textbf{55}, 1497 (1997). 
\bibitem{shukla-ppv}A. Shukla, Phys. Rev. B \textbf{65}, 125204 (2002). 
\bibitem{hng-ppv}A. Shukla, H. Ghosh, and S. Mazumdar, Phys. Rev. B \textbf{67}, 245203
(2003).
\bibitem{yu-su}J. Yu and W.P. Su, Phys. Rev. B \textbf{44}, 13315 (1991). 
\bibitem{orr}B.J Orr and J.F. Ward, Mol. Phys. \textbf{20}, 513 (1971). 
\bibitem{notation}The many-particle wave functions of half-filled conjugated polymers
such as PDPA exhibit particle-hole (charge conjugation) symmetry when
treated using the models such as the Hubbard model and the PPP model.
Therefore, for every configuration which contributes to its wave function,
its charge-conjugated (c.c.) counterpart will contribute in equal
measure. For example, if configuration $|H\rightarrow L+1\rangle$
occurs in the wave function with coefficient 0.69, configuration $|H-1\rightarrow L\rangle$
will also occur in the wave function with coefficient of the same
magnitude (although, the sign could be reversed). Thus, we will use
shorthand notation $|H\rightarrow L+1\rangle+c.c.$ (0.69) to denote
the contribution of these configurations to the wave function. Above,
$H/L$ refer to HOMO/LUMO orbitals.%
\end{thebibliography}
\end{document}